\documentclass[aps,prb,twocolumn]{revtex4-1} 
\usepackage{amsmath}
\usepackage{amssymb} 
\usepackage{graphicx} 
\usepackage{multirow}
\usepackage{amsfonts}
\providecommand{\U}[1]{\protect\rule{.1in}{.1in}} 

\begin{document}
\title{First-principles evidence of Mn moment canting in hole-doped
Ba$_{1-2x}$K$_{2x}$Mn$_{2}$As$_{2}$} 
\author{J. K. Glasbrenner} 
\affiliation{Code 6393, Naval Research Laboratory, Washington, DC 20375, USA} 
\author{I. I. Mazin}
\affiliation{Code 6393, Naval Research Laboratory, Washington, DC 20375, USA}
\date{\today}

\begin{abstract}
The compound BaFe$_{2}$As$_{2}$ is the proptotypical example of the 122 family
of high-$T_{c}$ Fe-based superconductors that crystallize in the
ThCr$_{2}$Si$_{2}$ structure. Isostructural compounds can be formed by replacing Fe 
with another transition metal; using Mn produces the material
BaMn$_{2}$As$_{2}$, which unlike its Fe-based cousin has an insulating ground
state with a large magnetic moment of $3.9 \mu_{B}$ and G-type
antiferromagnetic order. Despite its lack of superconductivity, the material is
interesting in its own right. Recent experimental studies have shown that
hole-doping the compound by substituting K for Ba leads to metallic behavior and
a spontaneous, weak, in-plane magnetization, which was attributed to the holes
fully polarizing independent of the Mn moments, producing half-metallic
behavior. However the observed in-plane magnetization can also be understood as
a small canting of the Mn moments. Using density functional theory, we demonstrate
 that a Mn moment canting occurs upon
hole-doping the compound. We argue that this is due to the competition between
the super- and double exchange interactions, which we support using a simple
tight-binding model of the superexchange-double exchange interaction and the
Andersen Force Theorem. Our calculations also
rule out an in-plane polarization of As holes as an explanation for the
in-plane magnetization.
\end{abstract}
\maketitle

\emph{Introduction$-$}The discovery of the high-$T_{c}$ Fe-based superconductors
in 2008 induced a flurry of interest as researchers worked to understand the
role of magnetism in the pairing mechanism and superconducting 
state.\cite{kamihara,HKM,mazin1,johnston,lumsdenchristianson,kordyuk,scalapino}
Much like the cuprates, the parent compounds of the Fe-based superconductors are 
magnetically ordered in the ground state, although the similarity diverges from there: 
the ground state of the parent compounds of the Fe-based superconductors is metallic 
with long-range magnetic correlations while in the cuprates the ground state is 
insulating with strong, local electronic correlations.\cite{mazin1} The Fe-based superconductors 
can be divided into different structural classes, including the 122 class which crystallizes 
into the ThCr$_{2}$Si$_{2}$ structure (space group I4/mmm). A protypical example of the 122 
class is BaFe$_{2}$As$_{2}$.\cite{huang,rotter} There has been much interest in studying 
other materials isostructural to BaFe$_{2}$As$_{2}$, such as replacing As with P
 or Se, or Fe with another transition metal, such as Co, Ni, Ru, or Mn.

BaMn$_{2}$As$_{2}$ is not the parent compound of any known superconductor, but
it is interesting in its own right. Unlike its cousin BaFe$_{2}$As$_{2}$, which
has a metallic ground state with stripe antiferromagnetic (AF) order and Fe moments of
$\sim0.9\mu_{B}$,\cite{huang} the ground state of BaMn$_{2}$As$_{2}$ is
insulating with G-type antiferromagnetic (G-AF) order. The Mn atoms have moments of
$\sim3.9\mu_{B}$,\cite{singhjohnston,singhgoldman,an} aligned along
the crystallographic $c-$axis. Metallic behavior can be induced 
through the application of pressure\cite{satya} or through doping, and
successful hole-doping was achieved by substituting Ba with
K.\cite{bao,pandey,pandeyjohnston} An ionic count suggests that Mn is divalent
and in the high-spin state, such that its mean field moment would be 5$\mu_{B},$
which is reduced by hybridization and fluctuations to $3.9 \mu_{B}$. The material
is a small band gap semiconductor with an intrinsic activation energy of $0.03$ eV,
 as inferred from electrical resistivity measurements.\cite{singhjohnston}

In heavily hole-doped samples a weak ferromagnetic (FM)
magnetization developes along an in-plane direction in
Ba$_{1-2x}$K$_{2x}$Mn$_2$As$_2$.\cite{bao,pandeyjohnston} When $x = 0.2$,
the measured FM magnetization was 0.45 $\mu_{B}$/f.u., close to the
number of introduced holes, so a novel magnetic state was speculated in which
the localized Mn moments remained G-AF ordered along the $c-$axis while the
mobile holes are polarized in the $ab$ plane.\cite{pandeyjohnston} The authors of 
Ref.~\onlinecite{pandeyjohnston} argued the hole polarization was
half-metallic, implying that if the density of states (DOS) is projected onto
the in-plane magnetization direction it will be metallic in one spin direction and
approximately insulating in the other.

The proposal of a novel state of two separate magnetic systems with localized
Mn moments and mobile holes in the perpendicular direction is somewhat
counterintuitive, as such a state is not well defined microscopically.
Indeed, the introduced holes in the Mn-As planes can
either be Mn holes or As holes (or a combination of the two). In the former case
the same electrons that form the local moments will also form Mn bands that host
 the mobile holes, but these electrons are subject to a strong Hund's rule
coupling and cannot form mutually orthogonal magnetic moments.
Since in this case one cannot distinguish
between the electrons forming local moments and mobile carriers when hole-doping, the
only way to implement the idea of mobile carriers promoting FM order is by
introducing canting, as in the case of the classical double
exchange.\cite{zener,andersonhasegawa}

In the latter case the mobile carriers are different (As holes) and can be polarized in a different direction. This would
require the DOS near the top of the valence band to be mostly As. In this
situation the problem is mathematically similar to the well-known case of
Co-doped FeS$_{2},$ where an analytical treatment predicts that the system may be
half-metallic or non-magnetic depending on the effective mass
and Stoner parameter $I$.\cite{mazin3} Despite this possibility, we will show below that As
polarization can be ruled out both numerically and analytically.

The main argument in Ref.~\onlinecite{pandeyjohnston} against Mn moments canting
was based on the lattice symmetry. These authors correctly point out that the
Dzyaloshinskii-Moriya interaction,\cite{moriya,dzayloshinskii} a common source
of such canting, is excluded here because the local Mn environment is symmetric
with respect to inversion about the Mn site. However, the Dzyaloshinskii-Moriya
interaction is not the only known source for such noncollinearity. As mentioned
above, the double exchange mechanism\cite{zener,andersonhasegawa} is also well
known for generating canting in metallic AF systems due to the competition
between the superexchange, which favors AF alignment, and
the tendency for mobile carriers to maximally delocalize, which favors
FM alignment.

The authors of Ref.~\onlinecite{pandeyjohnston} also analyze their NMR spectra in comparison with x-ray and
magnetic neutron diffraction, and conclude that a canting of Mn moments is
unlikely. However the authors appreciate that this is an indirect and involved
argument and considered it as secondary to their symmetry argument which, as
explained above, is not valid.

Correspondingly, we consider it an open question as to whether or not canting is
present in hole-doped Ba$_{1-2x}$K$_{2x}$Mn$_2$As$_2$, and in the following we will address it
using first principles calculations. We conclude that the system is canted and
the mechanism for that is the double exchange between the mobile holes and the
localized spins. We find that the DOS is not half-metallic when projected onto the
magnetization, contrary to the expectations of
Ref.~\onlinecite{pandeyjohnston}. Analytical considerations using parameters
derived from our first principles calculations give further support for our
conclusions.

\emph{Qualitative Considerations$-$}Let us first consider a hypothesis of
spin-polarized As holes. Inducing FM behavior upon doping would be
the result of a competition between the kinetic energy and the Stoner (Hund)
interaction $-I m^2/4$,\cite{mazin3} with the Stoner parameter $I_{As} = 1$ eV
for As. For $x = 0.2$ holes/As the system is heavily doped and can be
approximately thought of as a metal, in which case the Stoner criterion
$I_{\text{As}} N(0) > 1$ is appropriate. We calculated the DOS for collinear,
undoped BaMn$_2$As$_2$, attributing the full DOS to the As atoms, and shifted
the Fermi energy to simulate a doping level of 0.2 holes/As, finding $N(0) =
0.54$ spin$^{-1}$ eV$^{-1}$ As$^{-1}$ (see the next section for methods). The Stoner criterion is
therefore not satisfied and in-plane FM polarization would not be supported.
Later in this paper we will use the full DOS to compare $-I m^2/4$ against the
changes in the one-electron energies to confirm that the FM polarization of As holes is
not supported.

Now let us consider the double exchange scenario in which the carriers are Mn
holes and must, by virtue of the Hund's rule, be parallel to the local
moments. The condition for double exchange is $J_{H}\gg
t$,\cite{zener,andersonhasegawa} where $t$ is the one electron hopping
amplitude. Since the Hund's coupling $J_H$ in 3d metals is strong ($0.7-0.9$
eV) this condition is easily satisfied in BaMn$_{2}$As$_{2}$. Double exchange
requires mobile carriers and thus in BaMn$_{2}$As$_{2}$ would only emerge upon
doping. The mobile carriers can delocalize and lower their kinetic energy
if they are moving on a uniform FM background, and this preference for
ferromagnetic ordering must compete with superexchange which is responsible for
the observed G-AF order. It is known that for strong superexchange this
competition results in a canted state of angle
$\theta$,\cite{degennes,khomskii} where $\theta$ is the angle between the two antiparallel Mn moments ($180^{\circ}$ is G-AF order), and in a single-orbital tight-binding
approximation the explicit form for the canting angle is 
\begin{align}
\label{eq-degennes} \cos\left(  \frac{\theta}{2}\right)  =\frac{tx}{4Jm^{2}},
\end{align}
where $x$ is the doping per Mn, $J$ is the superexchange parameter,
and $m$ is the local Mn moment. Substituting typical values for $t$ and $J$ (later
we will present accurate calculations of these quantities), $t\sim200$ meV,
$J m^{2}\sim500$ meV, and the experimental moment $m = 3.9 \mu_B$, we obtain $\theta \sim 177.7^{\circ}$, implying that each Mn moment cants by $1.15^{\circ}$ and hence
$M_{FM} = 2 m \sin(1.15^{\circ}) = 0.16\mu_{B}/$f.u. This is on the right order, although
about a factor of three too small compared to the experiment.

Now we will present accurate calculations of the above quantities using
first-principles density functional theory.

\begin{figure*} \begin{center}
\includegraphics[width=0.90\textwidth]{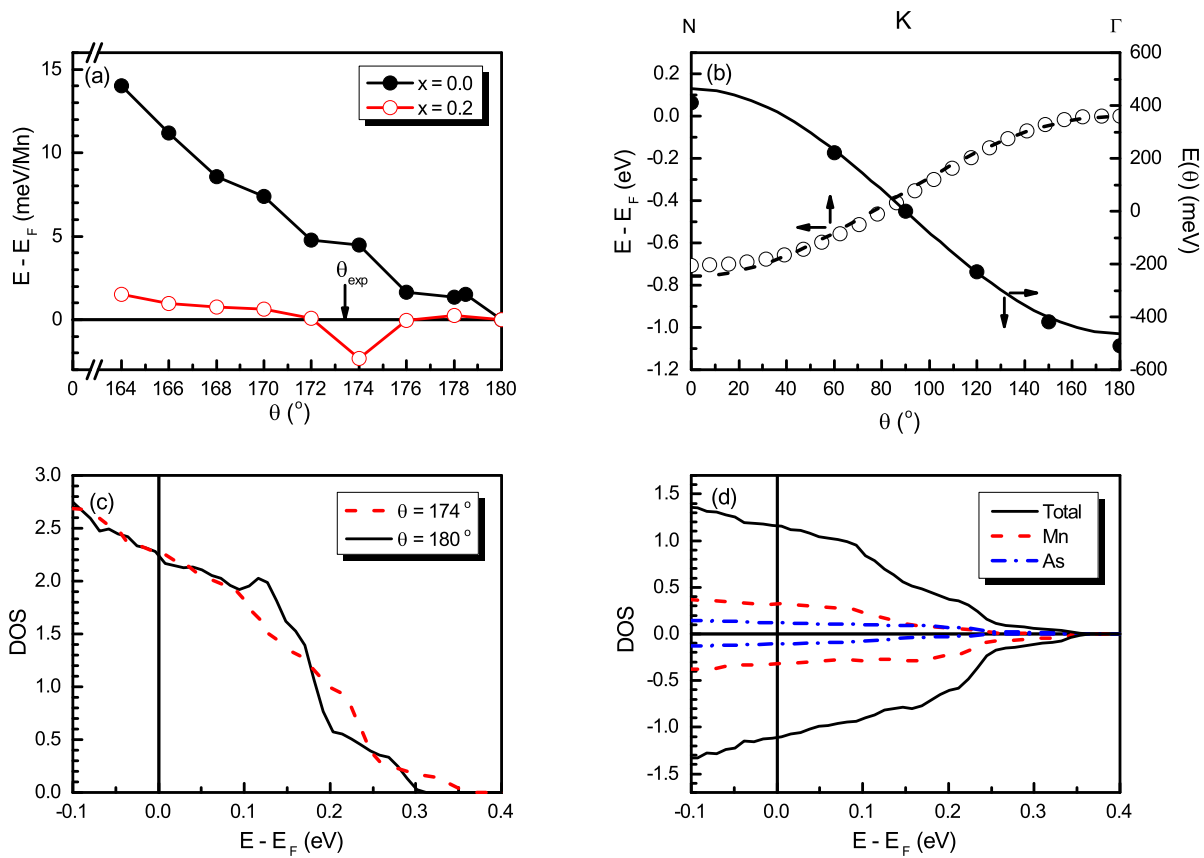} \end{center} \caption{(a)
The energy as a function of the angle $\theta$. The black curve is the undoped
($x=0.0$) case and the red curve is the doped ($x=0.2$) case. The black arrow references $\theta_{\text{exp}}$.\cite{pandeyjohnston} (b) \emph{Left and
bottom axes}: The closed black circles are the energy dependence of
undoped BaMn$_{2}$As$_{2}$ as a function of the relative angle between the Mn moments.
The black line is the fit of $J m^{2} \cos\theta$. \emph{Right and top axes}:
The open circles are the top valence band of BaMn$_{2}$As$_{2}$ along the
$N-\Gamma$ symmetry line. The dashed line is the nearest neighbor tight binding fit.
 (c) The total DOS for $x=0.2$ doping near
the Fermi energy for collinear (black line) and canted (dashed red line)
systems. (d) The spin-resolved As (blue dash-dot lines) and Mn (red dashed
lines) partial DOS and the spin-resolved total DOS (solid black lines) projected
along the direction of the in-plane magnetization.} \label{pic-results}
\end{figure*}

\emph{Computational Methods$-$}To perform our calculations, we used noncollinear
density functional theory (DFT) with the Perdew-Burke-Ernzerhof generalized gradient
 approximation\cite{pbe} to solve the electronic structure of
Ba$_{1-2x}$K$_{2x}$Mn$_{2}$As$_{2}$ using the full potential linear augmented
planewave code ELK\cite{elk} and PAW potentials as implemented in
VASP.\cite{vasp1,vasp2,paw1,paw2} As mentioned,
BaMn$_{2}$As$_{2}$ belongs to the space group I4/mmm, and we used lattice
parameters $a=4.16570 \text{ \AA }$ and $c=13.52110 \text{ \AA }$ and the
optimized
internal parameter for As $z_{\text{As}}$=0.358. Good convergence was
achieved with a $12 \times12 \times11$ k-point mesh and including 30 empty
eigenstates per atom per spin in the calculation. Doping was included in the
virtual crystal approximation (VCA) in ELK, where the Ba atom was replaced with
a fictitious atom of fractional charge, and by direct atomic substitution of K
for Ba in VASP.

\emph{Results$-$}All reported results were calculated using ELK unless otherwise noted. We confirmed that the ground state of BaMn$_{2}$As$_{2}$ is
G-AF. The calculated Mn
moments are $3.64 \mu_{B}$ within a muffin-tin radius of 1.259 {\AA}, in
reasonable agreement with experiment and previous calculations.\cite{an} We found an indirect band gap of $0.2$ eV, also in agreement with previous calculations.\cite{an} 

To study if canting can be stabilized, we
used the fixed-spin moment method to rotate the Mn moments in the $xz$ plane and
calculated the energy for several different values of $\theta$.
In this procedure the moment direction was constrained and the moment amplitudes
were allowed to relax. For our calculations we chose a VCA doping level of
$x=0.2$, corresponding to the hole doping level reported in
Ref.~\onlinecite{pandeyjohnston}. The results of these calculations are depicted in
Fig.~\ref{pic-results}(a).

The results in Fig.~\ref{pic-results}(a) show that the undoped system does not
exhibit canting, as is expected for an insulator without mobile carriers. For
$x=0.2$ the canting angle is predicted to be $\theta \approx 174^{\circ}$, or $3^{\circ}$ per Mn moment, in
excellent agreement with the angle defined in Ref.~\onlinecite{pandeyjohnston}
from the ratio of the experimentally measured FM and AF moments,
$\theta_{\exp}=2 \cos^{-1}(0.45/7.8)=173.4^{\circ}$ (shown as the black arrow in Fig.~\ref{pic-results}(a)). One can also see how doping
leads to strong cancellation between the super- and double exchange terms, as
even at $\theta = 164^{\circ}$ the energy difference with the collinear state is 1.5
meV/Mn, in contrast to 14.0 meV/Mn for the undoped case.

The energy scales for canting are on the order of a
couple meV as is seen in Fig.~\ref{pic-results}(a), which leads to difficulties
when trying to calculate the canting angle self-consistently using either VASP
and ELK. When using VASP and replacing 50\% of Ba atoms with K to simulate $x=0.25$,
we stabilized canting solutions with $\theta = 174^{\circ}$ and energy
$E(\theta)-E(0)=-0.8$ meV/Mn. However, depending on the canting angle used to
initialize the calculation, sometimes VASP relaxed to a larger canting angle.
The undoped compound always converged to the collinear solution. We ran into
similar problems with ELK. At $x=0.2$ we were able to converge to two different
solutions, one with $\theta \approx 176^{\circ}$ and energy $-0.25$ meV/Mn and the
other with $\theta \approx 168^{\circ}$ with energy $-0.06$ meV/Mn. It is clear
that the energy landscape is so complex that finding the global minimum is
difficult, and self-consistent calculations of the
canting moment are less reliable than the fixed-spin moment calculations. Apparently,
 the system does exhibit a tendency to cant, but the exact degree
of canting is difficult to determine.

We now get back to Eq.~\ref{eq-degennes} and determine its parameters from our
calculations in ELK. In Fig.~\ref{pic-results}(b) we calculated the energy of undoped 
BaMn$_2$As$_2$ as a function
of the relative angle between the two magnetic moments and fitted it to $E = J
m^{2} \cos \theta$, finding $J m^{2} = 463 \text{ meV}$. We also calculated the
band structure and fitted the
top valence band to the nearest neighbor tight binding model, also shown in Fig.~\ref{pic-results}(b).
The fit yielded the hopping amplitude $t=190\text{
meV}$. Using Eq.~\ref{eq-degennes} we find that $\theta=$177.7$^{\circ}$ for $x=0.2$,
in agreement with our previous rough estimate. This prediction is off by a factor of
2.6 when compared with the result of Fig.~\ref{pic-results}(a), which is reasonable given
 the simplicity of the model.

To address the microscopic origin of the canting observed in our DFT
calculations, it is instructive to compare the DOS for
$x=0.2$ in the VCA for both the uncanted case and
the canted case of $\theta = 174^{\circ}$, see Fig.~\ref{pic-results}(c). The
gain in kinetic energy from allowing the electrons to delocalize upon canting
can be estimated by using the Andersen Force Theorem\cite{AFT1,AFT2} and
calculating the change in the one-electron energy of the uncanted and canted
systems. Strictly speaking, the Force Theorem requires taking the same
charge and spin density for both cases; in the canted case the self-consistent
uncanted potential is rotated within each muffin-tin sphere by $3^{\circ}$ and
the DOS is generated non-self-consistently. This is not possible in ELK, so we
used the self-consistent canted DOS as a proxy assuming that the main changes in
DOS are due to canting and not by changing the spin density (indeed, the
calculated magentic moment is essentially the same $3.5603 \mu_B$ vs. $3.5606
\mu_B$). Applying the Force Theorem we can then approximate the total energy
change as the change in one-electron energy and the magnetic energy. The former
can be computed by integrating the DOS as $\int_{occ}E N(E)dE$ or, equivalently,
as $-\int_{unocc}E N(E)dE$ and normalizing the computed integral by the number of electrons or holes. The change in kinetic energy can be visualized as the broadening of the
unoccupied part of the valence band which results in an upshift of the center of
gravity. Using the DOS in Fig.~\ref{pic-results}(c), we find $\Delta
E_{\text{kin}} =$ 6.6 meV. The corresponding loss of the exchange energy is
$J m^2 (1+\cos\theta)$, and using $\theta=174.0^{\circ}$ gives us 2.5 meV.
 The energy gain in the one-electron energy is
about 2.6 times larger than the energy loss from the exchange interaction. This
indicates that canting is favored, but as in the case of relaxing the canting
angle self-consistently, the energy scales are quite small.

We now determine whether half-metallic behavior is possible in hole-doped Ba$_{1-2x}$K$_{2x}$Mn$_2$As$_2$, as argued in Ref.~\onlinecite{pandeyjohnston}. First we check whether a spin channel becomes approximately insulating when the Mn moments are canted. The
partial DOS for Mn and As along with the total DOS is projected along the
direction of the in-plane magnetization for the canted angle of $\theta =
174^{\circ}$ in Fig.~\ref{pic-results}(d). There is no evidence for half-metallic
behavior at the Fermi energy; the DOS is that of a weak ferromagnet. It should
be noted that the partial densities of states of As and Mn at around $E - E_{F}
\approx 0.25$ eV each become nearly half-metallic, although the polarization
directions of the two atoms oppose each other. This suggests that the emergence
of half-metallic behavior upon canting is possible, though it is not realized in
this system.

Finally we return to the scenario of polarized As holes. As before in our qualitative consideration we calculated the
DOS for collinear, undoped BaMn$_2$As$_2$ and shifted the Fermi energy to
simulate $x = 0.2$. Assuming that the full DOS can be attributed
to the As atoms, we then manually polarize the DOS and calculate $-I m^2/4$ and
compare it with the changes in the kinetic energy. We find that
the polarization of As holes is never favored. For full polarization, the gain
in Stoner energy is 10 meV while the kinetic energy loss of 150 meV is an order
of magnitude larger, so half-metallic polarization is very unfavorable. Of
course, in the actual hole-doped system the character of the carriers at the
Fermi energy is a combination of Mn and As states, with about three times
more Mn-like carriers than As-like carriers, as seen in
Fig.~\ref{pic-results}(d), so it is even more unlikely that the As holes could
polarize.

\emph{Conclusion$-$}We theoretically investigated hole-doped Ba$_{1-2x}$K$_{2x}%
$Mn$_{2}$As$_{2}$, in which weak ferromagnetism was discovered experimentally
and attributed to two groups of carriers, local electrons with spins aligned
along $c$-axis, and mobile holes fully polarized in the $ab$ plane. Our
first-principles calculations quantitatively reproduce the observed weak
ferromagnetism, yet the microscopic physics is better described by a canting of
Mn moments induced by double exchange. This conclusion is supported by our
numerical calculations and analytical analysis.

\bibliography{bma}

\end{document}